\documentclass[aps,prl,showpacs,twocolumn,superscriptaddress,letterpaper]{revtex4}

\usepackage{graphicx}
\usepackage{dcolumn}
\usepackage{bm}
\usepackage{amsmath, amssymb}%
\usepackage{epstopdf}
\usepackage{color}

\newcommand{\eq}[1]{Eq.~(\ref{#1})}

\begin{document}



\title{
Correlated  Fermion Pairs in Nuclei and Ultracold Atomic Gases}

\newcommand*{\TAU }{Tel Aviv University, Tel Aviv 69978, Israel}
\newcommand*{\TAUindex}{1}
\affiliation{\TAU} 
\newcommand*{\ODU }{ Old Dominion University, Norfolk, VA 23529,
USA} 
\newcommand*{\ODUindex}{2}
\affiliation{\ODU } 
\newcommand*{\UW}{University of Washington, Seattle, WA 98195-1560, USA}
\newcommand*{\UWindex}{3}
\affiliation{\UW}
\newcommand*{\FIU}{Florida International University, Miami, FL
  33199, USA}
\newcommand*{\FIUindex}{4}
\affiliation{\FIU}
\newcommand*{\TECHNION}{Department of Physics, Technion – Israel Institute of Technology, Haifa 32000, Israel}
\newcommand*{\TECHNIONindex}{5}

\affiliation{\TECHNION}

  
\author{O. Hen}
     \email[Contact Author \ ]{or.chen@mail.huji.ac.il}
     \affiliation{\TAU}
\author{L.B.~Weinstein}
     \affiliation{\ODU}
\author{E. Piasetzky}
     \affiliation{\TAU}
\author{G.A. Miller}
     \affiliation{\UW}
\author{M.M. Sargsian}
     \affiliation{\FIU}
\author{Y. Sagi}
     \affiliation{\TECHNION}

\date{\today}

\begin{abstract}

{\bf Background:} The high momentum distribution of atoms in two spin-state ultra-cold atomic gases 
with strong short-range interactions between atoms with different spins, which can be described 
using Tan's contact, are  dominated by short range pairs of different fermions and 
decreases as $k^{-4}$.   In atomic nuclei the momentum distribution of nucleons above the Fermi momentum
($k>k_F \approx 250$ Mev/c) is also dominated by short range
correlated different-fermion (neutron-proton) pairs.  

{\bf Purpose:} Compare high-momentum unlike-fermion momentum
distributions in atomic and nuclear systems.

{\bf Methods:} We show that, for $k>k_F$ MeV/c, nuclear
momentum distributions are proportional to that of the deuteron.  We
then examine the deuteron momentum distributions derived
from  a wide variety of  modern nucleon-nucleon potentials that are consistent
with $NN$-scattering data.  

{\bf Results:} The high momentum tail of the deuteron momentum distribution,
and hence of the nuclear momentum distributions
appears to decrease as $k^{-4}$.     This behavior is shown to arise from the effects of the tensor part of the nucleon-nucleon potential.
  In addition, when the dimensionless interaction strength for  the atomic system
is chosen to be similar to that of atomic nuclei, the probability for finding a short range 
different-fermion pair in both systems is the same. 

{\bf Conclusions:} Although nuclei do not satisfy all of the
conditions for Tan's contact, the observed similarity of the magnitude and
$k^{-4}$ shape of nuclear and atomic momentum distributions 
is remarkable because these systems differ 
by about $20$ orders of magnitude in density. 
This similarity may lead to a greater
understanding of nuclei and the density dependence of nuclear systems.

\end{abstract}


\maketitle

%


\noindent{\bf Introduction:} 
Interacting many-body Fermionic systems are abundant in nature.  In
non-interacting Fermi systems at zero temperature, the maximum
momentum of any Fermion in the system is the Fermi momentum, $k_F$.
Independent Fermions moving in a mean field potential have only a
small probability to have $k > k_F$. However, an additional
short-range interaction between fermions creates a significant
high-momentum tail.  In this work we discuss two very different
systems each composed of two dominant kinds of fermions: protons and
neutrons in atomic nuclei and two spin-state ultra-cold atomic
gases. While these systems differ by more than 20 orders of magnitude
in density, and the fermion-fermion interactions are very different,
both exhibit a strong short-range interaction between unlike fermions
creating short range correlated (SRC) pairs of unlike fermions that
dominate the high momentum tail.

The momentum distribution of a dilute two-component atomic Fermi gas with
contact interactions is known to exhibit a $C/k^4$ tail for $k>k_F$, where $C$ is
the contact as defined by
Tan~\cite{Tan08a,Tan08b,Tan08c,Braaten12,Stewart10,Kuhnle10,partridge05,werner09,schirotzek10,sagi12,gandolfi11,gandolfi13}.
The value of $C$ depends on the strength of interaction
between the two components, as parametrized by $a$, the scattering
length. Here we will show that although nuclei do not fulfill the
stringent conditions of Tan's relations, their momentum distribution
is remarkably similar to that of ultracold Fermi gases with the same
dimensionless interaction strength $(k_Fa)^{-1}$. The similarity is in
both its functional scaling and the spectral weight of the tail.

While this remarkable similarity may be accidental, it is plausible
that Fermi systems with a complicated non-contact interaction may
still posses universal properties on scales much larger than the scale
of the interaction. This approach might lead to greater insight into
nuclear pair-correlations as well as the behavior of density dependence of nuclear systems. 

This paper is structured as follows: we review our knowledge of
nucleon-nucleon pair correlations in nuclei, emphasizing that (1) the
momentum distribution of nucleons in nuclei at $k>k_F$ is dominated by
proton-neutron ($np$) pairs and (2) the momentum distribution of
nucleons in medium to heavy nuclei is proportional to that of the
deuteron at high momenta.  We then show that (3) the momentum
distribution of nucleons in the deuteron and hence in all nuclei
decreases approximately as $k^{-4}$ at high momenta, which (4) can be
understood from the short distance structure of correlations.  This
$k^{-4}$ distribution is (5) the same momentum
distribution as for the previously measured atoms in ultra-cold
two-spin-state atomic gases with a contact interaction.  We also show that (6) the pair
correlations probability for unlike fermions (i.e., the magnitude of
the momentum distribution at high momentum) is the same for nuclei and
for atomic systems when the dimensionless interaction strength of the
atomic system is chosen to be the same as for nuclei.  We then explore (7)
the applicability of the conditions of Tan's theory to atomic nuclei.
We discuss soft and hard nucleon-nucleon interactions in the Appendix.

Previous papers explored the nuclear momentum distribution as well as
relationships between atomic and nuclear systems.  Amado and Woloshyn
\cite{amado76} showed that the nuclear momentum distribution
$n(k)\propto (k^{-2} v(k))^2$ where $ v(k)$ is the fourrier transform
of the nucleon-nucleon potential.  This decreases as $k^{-4}$ if $
v(k)$ is momentum-independent.  Sartor and Mahaux found a more
complicated form for the momentum distribution at $k>k_F$ for a dilute
Fermi gas \cite{sartor80,sartor82}, although their momentum
distribution also decreases approximately as $k^{-4}$ for large
momenta.  Studies of the $^3$He Fermi liquid  might also be relevant to this topic
\cite{mazzanti04}.  Carlson {\it et al.}  compared quantum
monte carlo approaches to neutron matter and atomic physics \cite{carlson12}.  \" Ozen
and Zinner \cite{ozen14} proposed creating a two-component cold Fermi
gas closely analogous to nuclear systems.  Zinner and Jensen
\cite{zinner08,zinner13} explored the differences and similarities
between nuclei and cold atomic gases.  They point out that
``As the contact parameters are expected to be universal,
they should be the same for a nuclear system in the limit of large
scattering length.''  This work builds on these studies and examines
quantitatively the connections and similarities between two-component
atomic and nuclear systems.  Our analysis is different than recent
work that relates the nuclear contact to the Levinger Constant \cite{Weiss14}. 

\noindent{\bf (1) Short Range Correlations in Nuclei:} 
Atomic nuclei are among the most common many-body Fermi-systems.
Analysis of electron-nucleus scattering~\cite{moniz71} confirmed that
medium and heavy nuclei, with atomic weight $A\ge12$, exhibit the
properties of a degenerate system with a characteristic Fermi
momentum, $k_F \approx 250$ MeV/c.  However, experiments also show
that nuclei are not completely described by the independent particle
approximation
and that, as expected~\cite{Bethe:1971xm,Frankfurt81}, two-particle correlations are a leading
correction \cite{egiyan03,egiyan06,frankfurt93,fomin12,piasetzky06,subedi08,korover14}.  Nuclei are composed of
protons and neutrons with up and down spins, which can create six
different types of nucleon pairs. However, isospin invariance reduces
the types of independent pairs to four: spin-singlet proton-proton
($pp$), neutron-neutron ($nn$), and proton-neutron ($pn$) pairs and
spin-triplet $pn$ pairs.  Isospin symmetry further implies that all
three types of spin-singlet pairs are similar to each other, reducing
the types of pairs to two: spin-singlet and spin-triplet.  These pairs
have either even or odd values of the orbital angular momentum $L$
according to the generalized Pauli principle, $(-1)^{L+S+T}=-1.$

Experiments show that short-range correlated nucleon-nucleon pairs
account for approximately all of the high momentum, $k>k_F$, nucleons
in nuclei and about 20--25\% of all the nucleons in nuclei
\cite{egiyan03,egiyan06,frankfurt93,fomin12,piasetzky06,subedi08,korover14}.
They also show that short-range $np$ pairs dominate over $pp$ pairs
with a ratio $np/pp=18\pm5$ \cite{piasetzky06,subedi08,korover14},
even in heavy asymmetric nuclei such as lead~\cite{hen14}.  As $np$
pairs include contributions from both spin-singlet and spin-triplet
pairs whereas $pp$ pairs are entirely spin-singlet, the observed
$np/pp$ ratio implies that spin-triplet $np$ pairs account for
$85\pm3$\% of all pairs with spin-singlet isospin-triplet $pp$, $nn$
and $np$ pairs contributing $5\pm1$\% each for a total of $15\pm3$\%
spin-singlet pairs.  This is due to the dominant tensor interaction
(which acts only in spin-triplet states) between nucleons at relative
momenta between 300 and 600
MeV/c~\cite{sargsian05,schiavilla07,wiringa14}.  Corrections due to
correlations amongst three nucleons or more are
small~\cite{Bethe:1971xm,egiyan06} and appear only for nucleon momenta
greater than about 800 MeV/c.

\noindent{\bf (2) Nuclear momentum distributions:}
Because of the observed dominance of $np$ pairs in SRCs we can use the
independent-pair approximation~\cite{deShalit} to write the momentum
density at $k>k_F$ for heavier nuclei as:
\begin{equation}
n_A(k)  = a_2(A)  n_d(k),
\label{eq:scaling}
\end{equation}
where $n_A(k)$ and $n_d(k)$ are the high momentum parts of the nucleon
momentum distribution for a nucleus of atomic number $A$ and deuterium
respectively and the factor $a_2(A )$ is independent of $k$ and is the
probability of finding a high momentum pair in nucleus $A$ relative to
deuterium.  

This simple picture was validated experimentally by measurements of
the ratios of per-nucleon inclusive electron scattering cross sections
for nuclei of atomic number $A$ relative to deuterium at four-momentum
transfer squared, $Q^2= \vec q\thinspace^2 - \omega^2 > 1.5$ GeV$^2$
and Bjorken scaling parameter $1.5 < x < 1.9$ where $x =
Q^2/2m\omega$, $\vec q$ and $\omega$ are the three-momentum and energy
transferred to the nucleus, and $m$ is the nucleon mass.  Cross
sections in this kinematic region are sensitive to the integral of the
nucleon momentum distribution from a threshold momentum to infinity
where $k_{thresh}=k_{thresh}(Q^2,x)$ depends on $x$ and
$Q^2$~\cite{Arrington:2011xs}.  These cross section ratios are
independent of $x$ for $1.5\le x\le 1.9$ and $1.5 \le Q^2 \le 3$
GeV$^2$~\cite{egiyan03,egiyan06,frankfurt93,fomin12}, showing that the
momentum distributions have similar shapes for approximately $k_F\le k
\lesssim 3k_F$ ($ 275\pm25 \le k \lesssim 700$ MeV/c) validating
Eq.~\ref{eq:scaling}. The value of the ratio gives the proportionality
constant for the different nuclei
\begin{equation}
\label{eq:a2scale}
a_2(A)=\frac{\sigma_A/A}{\sigma_d/2}.
\end{equation}

\noindent{\bf (3) Deuteron momentum distributions:}
Since the momentum distributions of all nuclei at high momentum are
proportional to that of the deuteron for about $k_F\le k\le 3k_F$, we
now examine the deuteron momentum distribution.  We will show that the
nucleon momentum distribution for deuterium, and hence for all nuclei,
decreases approximately as $k^{-4}$ for the momentum range $1.3\,k_F \le k \le
2.5\,k_F$.  (In anticipation of the coming discussion of heavy nuclei,
we use $k_F=250$ MeV/c, the typical Fermi momentum for medium and
heavy nuclei.)

In order to study the range of possible deuteron momentum
distributions, we considered ten modern nucleon-nucleon ($NN$)
potentials that are consistent with the nucleon-nucleon scattering
world data
set, the Nijmegen 1, 2, and
  3 \cite{Nijmegen94}, AV18 \cite{wiringa95}, CD Bonn \cite{CDBonn01},
  wjc1 and 2 \cite{wjc08}, IIB \cite{gross92}, n3lo500 and n3lo600
  \cite{Machleidt20111} nucleon-nucleon interactions.

The chiral effective field theory ($\chi$EFT) N3LO
potentials~\cite{Machleidt20111} have an explicit momentum cut-off of
the form $\exp[-(p/\Lambda)^n]$ where $n=4,6$ or 8 and $\Lambda=500$
or 600 MeV.  While we use the N3LO potentials with 500 and 600 MeV
cutoffs, the $\chi$EFT neutron-proton phase shifts differ dramatically
in some partial waves (especially at higher energy) as the cutoff is
varied from 0.7 to 1.5 GeV or as the expansion order is increased from N3LO to
N4LO \cite{entem15}.  In addition, ``the N2LO, N3LO, and N4LO
contributions are all about of the same size, thus raising some
concern about the convergence of the chiral expansion for the
$NN$-potential \cite{entem15b}.'' The N5LO contribution is much
smaller, indicating convergence \cite{entem15b}.  It is unclear how
these convergence issues affect the ability of N3LO potentials to
describe the high-momentum tail of the deuteron.  However, high
precision deuteron momentum distributions are not yet available for
higher order $\chi$EFT.

There is also some disagreement over the utility of bare interactions versus
soft phase-equivalent interactions.  This is discussed in detail in
the  Appendix.

\begin{figure} [htbp]
\includegraphics[width=3.5in]{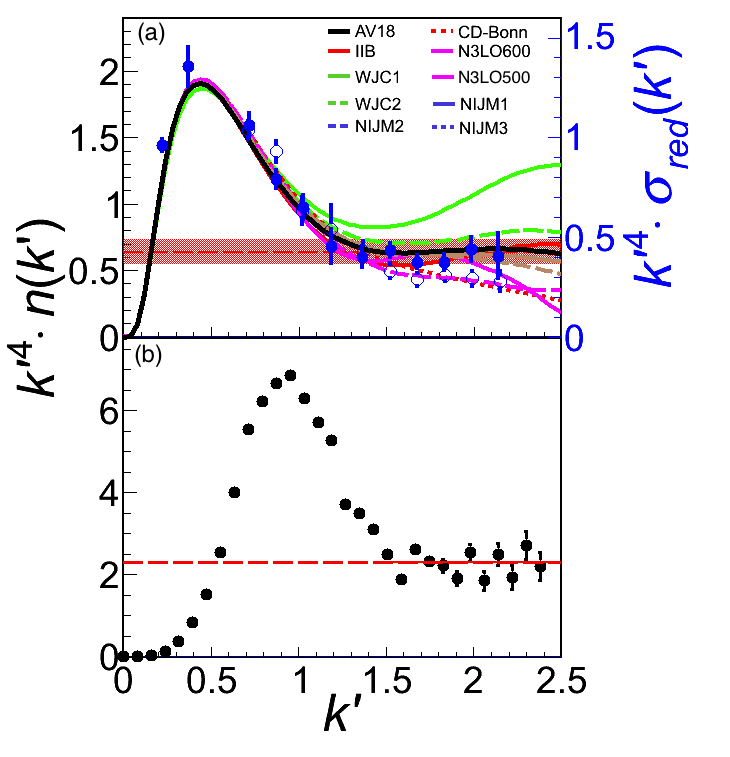}
\caption{\label{fig:k4nk} (color online) The scaled momentum
  distribution, $k'^4n(k')$ where $k'=k/k_F$, for deuteron (a) and
  atomic (b) systems.  (a) The curves show the scaled proton momentum
  distribution for the deuteron calculated from the Nijmegen 1, 2, and
  3 \cite{Nijmegen94}, AV18 \cite{wiringa95}, CD Bonn \cite{CDBonn01},
  wjc1 and 2 \cite{wjc08}, IIB \cite{gross92}, n3lo500 and n3lo600
  \cite{Machleidt20111} nucleon-nucleon interactions.  The dashed red
  line is the average of eight of the calculated momentum
  distributions for $k\ge 1.3k_F$.  The red band shows the $\pm$15\%
  uncertainty.  The points show the scaled reduced cross sections
  (using the right-hand $y$-axis), $k'^4\sigma_{red}(k')$, for
  electron-induced proton knockout from deuterium, $d(e,e'p)$, at
  $\theta_{nq}=35^\circ$ (filled circles) and at
  $\theta_{nq}=45^\circ$ (open circles)\cite{Boeglin:2011mt}.  The
  curves and points are plotted in units of $k_F=250$ MeV/c, the
  typical Fermi momentum for medium and heavy nuclei. This choice of
  $k_F$ affects the normalization but not the observed $k^{-4}$
  scaling.  (b) The points show the measured momentum distribution of
  $^{40}$K atoms in a symmetric two-spin state ultra-cold gas with a
  short-range interaction between the different spin-states
  \cite{Stewart10}.  The dimensionless interaction strength $(k_F
  a)^{-1}=-0.08\pm0.04$. The Fermi momentum is $k_F\approx 1.6$ eV/c.}
\end{figure}

The momentum distribution of a nucleon bound in deuterium, $n_d(k)$,
was calculated using each of the modern nucleon-nucleon potentials.
The proton and neutron momentum distributions in the deuteron are
equal, $n_p(k/k_F) = n_n(k/k_F)=n_d(k/k_F)$, and are normalized so
that 
\begin{equation}
\frac{1}{(2\pi)^3} \int_0^\infty n(k/k_F) d^3(k/k_F) = 1/2. 
\label{eq:norm}
\end{equation}

We can see the $k^{-4}$ dependence of the momentum distribution
clearly in Fig.~\ref{fig:k4nk}a, which shows the scaled dimensionless
momentum distribution, $(k/k_F)^4 n_d(k/k_F)$, for a nucleon bound in
deuterium for each of these potentials~\cite{Machleidt20111}.  We
observe $k^{-4}$ scaling in seven of the ten different realistic
models, all showing that the ratio
\begin{equation}
R_d = (k/k_F)^4  n_d(k/k_F) = 0.64\pm0.10  
\end{equation}
for $1.3 \le k/k_F \le 2.5$ is constant within about 15\% as shown by
the red dashed line and uncertainty band in Fig.~\ref{fig:k4nk}a.
Note that $k^{-4}$ changes by a factor of 14 in this range and even
the outlying potentials only differ by at most a factor of two from
the average.

 The experimental reduced $d(e,e'p)$ cross sections for $1.3\,k_F\le k
 \le 2.5\, k_F$ also appear to scale as $k^{-4}$ and provide more
 evidence for the scaling of the momentum distribution.
 Fig.~\ref{fig:k4nk}a also shows the measured $d(e,e'p)$ scaled
 reduced cross sections, $(k/k_F)^4 \sigma_{red}(k/k_F)$, for proton
 knockout by electron scattering from deuterium in two kinematics
 where the effects of rescattering of the knocked-out proton (final
 state interactions or FSI) are expected to be small
 \cite{Boeglin:2011mt}.  The two kinematics are for the angle between
 the undetected neutron and the momentum transfer,
 $\theta_{nq}=35^\circ$ and $45^\circ$.  If the electron interacts
 directly with an on-shell proton and the proton does not rescatter as
 it leaves the nucleus, then the reduced cross section equals the
 momentum distribution.  Corrections for these effects are model
 dependent and are on the order of 30--40\% (see
 Ref. \cite{Boeglin:2011mt} and references therein).  These effects
 should be significantly smaller for $\theta_{nq}=35^\circ$ than for
 $45^\circ$.  The momentum dependence of these effects should also be
 significantly smaller.

\begin{figure} [htbp]
\includegraphics[width=2.5in, height=2.5in]{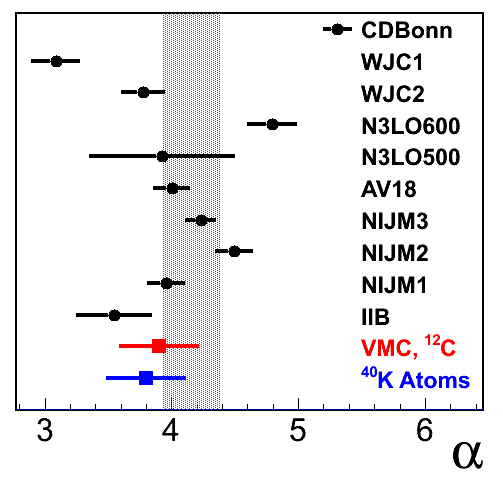}
\caption{\label{fig:k4_alpha} (color online)  The power $\alpha$ 
  obtained by fitting the momentum distribution of the nucleon in
  deuterium to the  form: $n_d(k)\propto k^{-\alpha}$ over $1.3k_F\le
  k \le 2.5 k_F$ for the nucleon-nucleon interactions of Fig.~\ref{fig:k4nk}.
The momentum distributions from n3lo500 and  n3lo600 are sharply regulated (forced to
decrease rapidly) at around 500 ($k= 2k_F$) and 600 MeV/c ($k/k_F=
2.4k_F$) respectively, so we restricted the upper limit of their fit
ranges to $2.0k_F$ and $2.4k_F$.
 The uncertainty of the nuclear momentum distribution exponent comes
 predominantly from varying the lower and upper bounds of the fitting
 range by $\pm0.1k_F$ and $\pm0.2k_F$ respectively.
 The red band is the average $\alpha$  ($\pm2\sigma$) obtained from the deuteron distributions, 
 excluding the two outlier wave-functions: CDBonn and WJC1.  Also
 shown is the result of the power-law fits  for the momentum
 distributions of a nucleon in $^{12}$C \cite{wiringa14} and an atom in an ultra-cold two-spin-state  $^{40}$K gas \cite{Stewart10}.  }
\end{figure}

We
fitted these momentum distributions by $n_d(k)\propto k^{-\alpha}$ for $1.3k_F
\le k \le 2.5k_F$ (except for N3LO500 and N3LO600 which we fit up to
their cutoffs of 500 MeV/c ($2k_F$) and 600 MeV/c ($2.4k_F$)
respectively).  We varied the lower and upper fitting bounds by
$\pm0.1k_F$ and $\pm0.2k_F$ respectively to determine the uncertainty
in the exponent $\alpha$ (see Fig.~\ref{fig:k4_alpha}).  We observe
$k^{-4}$ scaling in seven of the ten different realistic models of the
nucleon momentum distribution in deuterium.

This scaling behavior arises from the sum of the $S$ and $D$ wave
contributions to the density.  Due to the tensor interaction, the high
momentum tail is predominantly produced by $J= 1,\,S,D$-wave $np$
nucleon pairs ($T=0, S=1, L=0,2$ or $^3S_1-^3D_1$) \cite{vanhalst14}.
For larger momenta, $k>2.5k_F$, the momentum distribution falls more
rapidly with $k$. However, this accounts for less than 1\% of the
fermions in the system~\cite{egiyan06}.  The momentum distribution of
$pp$ pairs does not scale, since there is a minimum in the momentum
distribution at $k/k_F \approx 1.6$.

In agreement with the $np$-pair dominance model, exact calculations of
the $^{12}$C momentum distribution using the AV18 potential also show
$k^{-4}$ scaling \cite{wiringa14}.  Rios, Polls and Dickhoff
calculated the momentum distribution for infinite symmetric nuclear
matter using a self-consistent Green's function (SCGF) approach for
the AV18, CDBonn and N3LO500 interactions \cite{polls09,rios14}.  They
show that the N3LO500 potential fails to reproduce the measured
deuteron momentum distributions.  Their AV18 and CDBonn nuclear matter
momentum distributions decrease with approximately the same exponent
$\alpha$ as do the AV18 and CDBonn deuteron momentum distributions
shown in Fig. \ref{fig:k4_alpha}.  

Based on the inclusive $A(e,e')$ cross section ratios discussed above,
the $np$-pair dominance model, and the calculations of nuclear and
nuclear matter momentum distributions, we conclude that the momentum
distributions of all nuclei decrease approximately as $k^{-4}$.

{\bf (4) Understanding the $k^{-4}$ scaling}: This scaling should not
be surprising.  Colle et al. \cite{colle15} found that the nuclear
mass dependence of the number of SRC $pp$ and $pn$ pairs in nuclei can
be described by tensor operators acting on $NN$ pairs in a nodeless
relative $S$-state of the mean-field basis \cite{ryckebusch15}.  This
very short range behavior of the correlated part of the $NN$
interaction leads to $k^{-4}$ momentum dependence at high
momentum, as is shown next.

This $k^{-4}$ behavior
can  be understood to arise from the importance of the one pion
exchange (OPE) contribution to the tensor potential $V_T$, acting in
second order.  The Schoedinger equation for the spin-one two-nucleon
system, which involves $S$ and $D$ state components, can be expressed
as an equation involving the $S$ state only by using $(-B
-H_0)|\Psi_D\rangle =V_T|\Psi_S\rangle$, where $B$ is the binding
energy of the system and $H_0$ is the Hamiltonian excluding the tensor
potential. Thus one obtains an effective $S-$state potential: $V_{\rm
  00}=V_T( -B-H_0)^{-1}V_T$, where $V_T$ connects the $S$ and $D$
states.  The intermediate Hamiltonian $H_0$ is dominated by the
effects of the centrifugal barrier and can be approximated by the
kinetic energy operator.  This second-order term is large because it
contains an isospin factor $(\mathbf\tau_1\cdot\mathbf\tau_2)^2=9$, and
because $S_{12}^2=8-2S_{12}$.  Evaluation of the $S$-state potential,
neglecting the small effects of the central potential in the intermediate $D$-state,
yields
\begin{equation} V_{00}(k,k')\approx-M\frac{16f^4}{\mu^4 \pi^4}\int
  \frac{p^2pdp}{MB+p^2}I_{02}(k,p)I_{20}(p,k'),\label{v2}\end{equation}
where $M$ is the nucleon mass, $f^2\approx 0.08$ is the coupling
constant, $\mu$ is the pion mass, and $I_{LL'}$ are Fourier transforms
of the OPE tensor potential.
\[I_{02}(p,k)=I_{20}(p,k)={k^2Q_2(z)+p^2Q_0(z)\over 2pk}-Q_1(z),\]
with $z\equiv (p^2+k^2+\mu^2)/( 2pk),$ and $Q_i$ are Legendre
functions of the second kind. The important property is that
$\lim_{p\to\infty}
I_{02}(p,k)=1-(k^2+\mu^2)/p^2+\cdots$. 
Thus the integrand of \eq{v2} is dominated by large values of $p$ and
diverges unless there is a cutoff. This means that $V_{00}( k',k)$ is
approximately a constant, independent of $ k$ and $k'$. This is the
signature of a short ranged interaction.  As discussed in the
Introduction, this is the necessary and sufficient condition to obtain
an asymptotic two-nucleon wave function $\sim 1/k^2$ and a momentum
density $n(k) \sim 1/k^4$. Potentials that do not yield
  this behavior either have a very weak tensor force or a momentum
  cutoff at low momenta.
 

\noindent{\bf (5) Comparing the nuclear and atomic high momentum tails:}
Fig.~\ref{fig:k4nk}b shows the $k^{-4}$ scaling of $^{40}$K
 atoms. Note the remarkable similarity between the data depicted in Fig.~\ref{fig:k4nk}a and  
 ~\ref{fig:k4nk}b.
 The nuclear momentum distributions have the same $k^{-4}$ dependence
as the momentum distribution measured for two spin-state ultra-cold
$^{40}$K atoms of Ref.~\cite{Stewart10} and as Tan's predictions.




\noindent {\bf (6) Nuclear and atomic pair correlations probabilities:}
After establishing the similarity in $k^{-4}$ scaling of the momentum distribution tail of the nuclear and cold atoms systems, we now compare the spectral weight contained in these tails.
Similar to atomic gases, we define the normalized dimensionless
scaling coefficient per particle as
\begin{equation}
\frac{C}{k_F A}\equiv  (k/k_F)^4  n(k/k_F)
\label{eq:contact}
\end{equation}
at high momentum, where $A$ is the number of fermions in the system
and $n(k/k_F)$ is the dimensionless scaled fermion momentum distribution in units of $k_F$,
normalized according to Eq.~\ref{eq:norm}. $C/(k_F
A)$  is a measure of the per particle number of short-range correlated pairs.  For nuclei 
\begin{equation}
\label{eq:scaling_CA}
\frac{C}{k_F A}  = a_2(A) R_d,
\end{equation}
where $C/(k_F A)$, for nuclei,  is the sum of all four possible coefficients,
dominated by  spin-triplet $np$ pairs, and the ratios $a_2(A)$ are taken from \cite{fomin12}.    
See Table \ref{tab:contact}.

\begin{table} [htbp!]
\begin{center}
{\baselineskip 13pt
\begin{tabular}{|c|c|c|} \hline
Nucleus & $a_2(A)$ & $C/(k_F A)$ \\ \hline
$^{12}$C & $4.75\pm0.16$ & $3.04\pm0.49$ \\ \hline
$^{56}$Fe & $5.21\pm0.20$ & $3.33\pm0.54$ \\ \hline
$^{197}$Au & $5.16\pm0.22$ & $3.30\pm0.53$ \\ \hline
\end{tabular}
}
\end{center}
\caption{The scaling coefficient extracted for different nuclei.  $a_2(A)$ is the
  ratio of the per nucleon inclusive $(e,e')$ cross sections for
  nucleus $A$ relative to deuterium for $Q^2>1.5$ (GeV/c)$^2$ and $1.5
  \le x \le 1.9$ \cite{fomin12} (see Eq.~\ref{eq:scaling}).  $C$ is defined by Eq.~\ref{eq:scaling_CA}. 
}
\label{tab:contact}
\end{table}

\begin{figure} [tbp]
\includegraphics[width=3.2in]{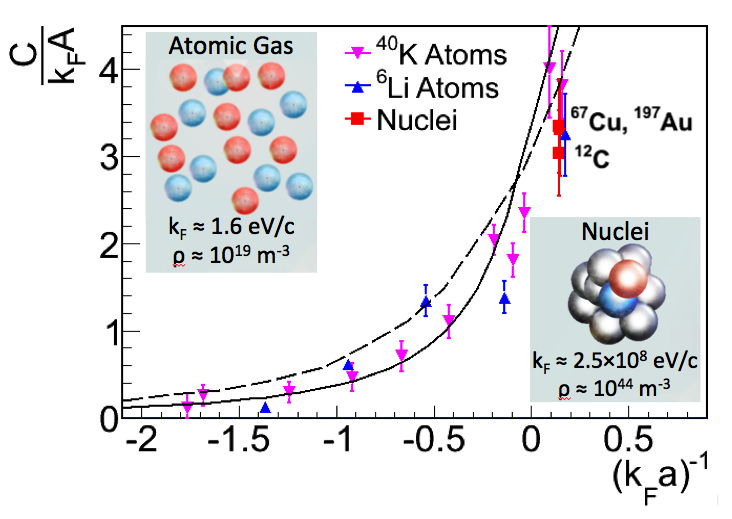}
\caption{\label{fig:contact} (color online) The magenta inverted and
  blue upright   triangles show the scaled contact plotted versus
  $(k_Fa)^{-1}$, the inverse of the product  of the scattering length and Fermi momentum, as 
  extracted from  measurements of ultra-cold two-spin state atomic systems at finite temperature
  \cite{Stewart10,Kuhnle10}.  The red squares show the equivalent
  coefficient extracted from atomic nuclei (see Table \ref{tab:contact}), which are essentialy at zero 
  temperature.   The dashed and solid lines show the theoretical predictions
  of Refs.~\cite{werner09} and \cite{Haussmann09} respectively for atomic systems at zero temperature.  }
\end{figure}

Fig.~\ref{fig:contact} shows the nuclear coefficients from Table
\ref{tab:contact} and the scaled atomic contact as extracted from
measurements of the momentum distribution of trapped two spin-state
mixtures of ultra-cold $^{40}$K \cite{Stewart10} and $^6$Li
\cite{Kuhnle10} atomic gases as a function of the dimensionless
interaction strength, $(ak_F)^{-1}$.  The ultra-cold atomic gas
measurements span a wide range of interaction strengths near unitary, in the BCS-BEC 
crossover regime.  In the nuclear case all medium and heavy nuclei are in the unitary 
regime where $\vert k_F a
\vert^{-1}\ll 1$, using the typical nuclear Fermi momentum, $k_F=250$
MeV/c $=1.27$ fm$^{-1}$ \cite{moniz71} and the $^3S_1$ neutron-proton
scattering length, $a=5.42$ fm \cite{Koester75}.  As can be seen, when
the dimensionless interaction strength is the same, the scaled atomic
contact and the nuclear coefficient agree remarkably well.  The integral of the $k^{-4}$ tail of the momentum density is about 0.2. Thus, each fermion has a
$\approx 20$\% probability of belonging to a high-momentum
different-fermion pair in both the atomic and nuclear systems.

This agreement between the shape and magnitude of the momentum distributions between such wildly disparate systems is remarkable.  We now look for the underlying reasons for this agreement.

\noindent {\bf (7) Possible Connections to Tan's Contact:}
Tan \cite{Tan08a,Tan08b,Tan08c} and later others
\cite{Braaten12}, showed that a short-range interaction between two
different Fermion types leads to a high-momentum tail that falls as
$k^{-4}$ (where $k$ is the fermion momentum), and derived a series of
universal relations that relate the contact (i.e. the number of
short-range correlated pairs) to various thermodynamic properties of
the system such as the total energy and pressure.
These were recently validated experimentally in ultra-cold two-spin 
states atomic gas systems~\cite{Stewart10,Kuhnle10}, see~\cite{Braaten12} for a review. 

Tan obtained relations for dilute systems with scattering length, $a$,
and the inter-fermion distance, $d$, which are much larger than the
range of the interaction, $r_0$: $a \gg r_0$ and $d \gg r_0$.  In such
systems the $k^{-4}$ scaling of the momentum distribution is only
expected for $ka\gg1$ and $k r_0 \ll1$.

As we have shown, atomic nuclei exhibit some of the same key
properties as cold atomic Fermi systems. They are characterized by a
Fermi momentum $k_F$, and have a strong short-range interaction
between (spin-triplet) unlike fermions. The nuclear momentum
distributions also fall as $k^{-4}$ for $300 \le k \le 600$ MeV/c.

However, unlike systems of atoms, atomic nuclei are self-bound.  The
nucleon-nucleon force provides both the long range interactions that
cause atomic nuclei to resemble Fermi gases and the short-range
interaction between fermions.  The binding interaction arises in part
from the iterated effects of the long-distance one-pion exchange
potential and has a range of about $r_0^{bind}\approx \hbar/(m_\pi c)\approx
1.4$ fm, where $m_\pi=140$ MeV/c$^2$ is the pion mass. The range of
the short-range part of the nucleon-nucleon interaction responsible
for the spin-triplet $pn$ pairs in the high momentum tail is less well
defined.  The $NN$-pairs in a nodeless relative $S$-state
\cite{colle15} are at much closer distances than typical nucleons.
Similarly, the second-order action of the tensor interaction described
in Eq. \ref{v2} also must have an effective range much shorter than
the long distance pion exchange that binds the nucleus.
Quantitatively, various tensor
correlation functions shown in \cite{vanhalst12} peak at an
internucleon distance of about 1 fm, so we will estimate that
$r_0\approx 1$ fm.
The typical distance between same-type nucleons
in nuclei is $d=(\rho_0/2)^{-1/3}\approx 2.3$ fm, where
$\rho_0\approx0.17$ nucleons/fm$^3$ is the saturation nuclear
density. The nucleon-nucleon scattering length in the $^3S_1$ channel
is 5.424 $\pm$ 0.003 fm \cite{Koester75}.

Therefore, for nuclei both the interaction length and the
inter-nucleon distance are greater, but not much greater, than the
range of the short-distance interaction (i.e. $a\approx 5.4
\thinspace\hbox{fm} > d\approx 2 \thinspace\hbox{fm} > r_0\approx 1$
fm).  Other required conditions for $1/k^4$ scaling are $k\gg
1/a\approx 40$ MeV/c and $k\ll 1/r_0 \approx 200$ MeV/c for
$r_0\approx 1$ fm.  As can be seen in Fig.~\ref{fig:k4nk}a, scaling
occurs for $300 \le k \le 600$ MeV/c, much greater than the lower
limit of 40 MeV/c.  However, $k$ is definitely not much less (or even less) than the upper
limit of 200 MeV/c.

The required kinematic conditions discussed above are sufficient, but
perhaps not necessary.  The scaling of quark distributions measured in
deep inelastic electron scattering was observed at momentum transfers
much below that expected.  This is referred to as `precocious scaling'
\cite{georgi76}.  Additionally the $1/k^4$ tail in ultracold Fermi
gases was experimentally observed to start at a much lower momentum
than predicted ~\cite{Hu:2010sx, Stewart10}.

 

Our extraction is different than a recent work by Weiss, Bazak, and
Barnea \cite{Weiss14} that relates the nuclear contact term to the
Levinger constant.  They attempt to extract it from
photodisintegration data, which are driven by the electric dipole
operator that operates on neutron-proton pairs.  Photodisintegration
is not a measure of the ground-state nucleon momentum density.
Furthermore their analysis is restricted to photon energies below 140
MeV, which corresponds to wavelengths $\lambda\ge 2\pi\hbar c/E = 8.8$
fm which sample the entire nucleus and are not short-range.  Also,
these photon energies correspond to nucleon momenta less than about
340 MeV/c ($\approx 1.35 k_F$).  This is below the $k^{-4}$ scaling
region shown in Fig. \ref{fig:k4nk} above.  Their average contact
(singlet plus triplet over two) equals our triplet contact alone, so
that their total contact is double ours and also double that of an
atomic system with the same value of $k_Fa$.

\noindent {\bf Summary:}
We have shown that the momentum
distribution of nucleons in nuclei for $k>k_F$ is dominated by
spin-triplet $pn$ pairs and falls approximately as $k^{-4}$.  This is very similar
to   the momentum distribution of  two
spin-state ultra-cold atomic gases with a strong short-range
interaction between atoms in the different spin states.  Remarkably,
despite a 20-order-of-magnitude difference in density,
when both systems have the same dimensionless interaction strength,
$(k_F a)^{-1}$, the magnitudes of the momentum distributions are also
equal, indicating that Fermions in the two systems have equal
probabilities to belong to correlated pairs.


This leads to the question of whether this
agreement between atomic and nuclear systems at remarkably different
length, energy and momentum scales is accidental or has a deeper
reason.  If the agreement has a deeper reason, then perhaps relations
like Tan's can be developed for atomic nuclei and a better extrapolation to 
supra-dense nuclear systems may be  possible.


\begin{acknowledgments}

 We thank  W. Boeglin, E. Braaten, W. Cosyn, L. Frankfurt,   D. Higinbotham, S. Moroz,
 J. Ryckebusch, R. Schiavilla, M. Strikman, J.W. Van Orden, and
 J. Watson  for many fruitful discussions. We particularly thank
 J.W. Van Orden for providing the deuteron momentum distributions.
 We also thank  D. Jin for the atomic data presented in Fig. 2. This work was partially supported by 
 the US Department of Energy under grants DE-FG02-97ER-41014, 
 DE-FG02-96ER-40960, DE-FG02-01ER-41172 and by the Israel Science Foundation.
\end{acknowledgments}

\section{Appendix: Soft and Hard $NN$ Interactions}

Nuclear theory must describe a broad
range of phenomena from low energies to high energies and from low
momentum transfer to high momentum transfer.  Therefore one must
contend with the fact that the general baryon-baryon interaction
includes matrix elements that connect low relative momenta to high
relative momenta. These terms can be handled by using so-called soft
$NN$ potentials and by generating soft phase-equivalent effective
interactions obtained from the bare interactions by means of unitary
transformations~\cite{bogner03,feldmeier98,roth10,bogner07,bogner10,wendt11,neff15}.
 
Calculations of low-energy and low-momentum processes are indeed
simplified by using soft interactions, and it is reasonable to obtain
such interactions using a unitary transformation.  A consistent
application of this idea involves transforming the Hamiltonian and all
other operators, especially including the currents that account for
interactions with external probes \cite{neff15}. Such transformation are known to
convert single-nucleon operators into multi-nucleon operators.  The
effect on long-range operators such as the radius or electromagnetic
transition operators is small~\cite{wendt11,neff15}.
However, the effect on short-range or high-momentum observables is
large~\cite{neff15}.

To posit the soft-interaction to be the fundamental bare interaction
is to deny the reality of high momentum transfer processes.
Therefore, the fundamental bare interaction must
allow high-momentum transfer processes.

In particular, consider that two-body densities in coordinate space
have a correlation ``hole" near $r=0$. By transforming only the
Hamiltonian and not the two-body density, these correlation holes
disappear~\cite{neff15}.  Similarly, these Hamiltonian-only
transformations dramatically reduce the high-momentum part of the
momentum density.

As an example, consider coherent neutrino-nuclear
interactions~\cite{scholberg06,akimov13,horowitz03}.  The neutrino
interacts weakly and the cross section is proportional to the square
of the elastic nuclear form factor. Following~\cite{horowitz03} we
note that the neutrino-nucleus elastic-scattering cross section
$d\sigma/d\Omega$ is \cite{freedman77,drukier84},
\begin{equation}
{d\sigma\over d\Omega} =   {G^2\over 16\pi^2} k^2 (1+{\rm cos}\theta)F^2(Q^2),
\end{equation}
for a neutrino of energy $k$ scattering at angle $\theta$, and $G$ is
the Fermi coupling constant.  The ground-state elastic form factor
$F(Q^2)$ at momentum transfer $Q$, $ Q^2=2k^2(1-{\rm cos}\theta), $ is
the matrix element of the single-nucleon operator $e^{i{\bf q}\cdot
  {\bf r}_i},$ for a nucleon $i$ weighted by the weak charge of the
proton or neutron.  One could ideally contemplate probing the
high-momentum components of nuclear wave functions using
neutrino-nuclear interactions. Now imagine that one wished to describe
the nuclear wave function using soft interactions. The necessary
unitary transformation would transform the simple single-nucleon operator
$e^{i{\bf q}\cdot {\bf r}_i},$ into a complicated multi-body operator,
which would ruin the simplicity of using the neutrino as a probe.

If one wishes to use simple probes to investigate high-momentum
aspects of nucleon structure, it is necessary to start with a theory
involving bare interactions, which are necessarily hard interactions.

\bibliography{eep}


\end{document}